\documentstyle[11pt,epsfig]{article}
\title{\bf Curvature corrections in DGP brane cosmology}
\author{K. Atazadeh\thanks{email: k-atazadeh@sbu.ac.ir}\,\ and H. R.
Sepangi\thanks{email: hr-sepangi@sbu.ac.ir}
\\ {\small Department of Physics, Shahid Beheshti University, Evin,
Tehran 19839, Iran}}
\begin{document}
\maketitle

\begin{abstract}
We consider a DGP inspired brane scenario where the action on the
brane is augmented by a function of the Ricci scalar, ${\cal
L}(R)$. The cosmological implications that such a scenario entails
are examined for $R^{n}$ and shown to be consistent with a
universe expanding with power-law acceleration. It is shown that
two classes of solutions exist for the usual FRW metric and small
Hubble radii. When the Hubble radius becomes larger, we either
have a transition to a fully 5D regime or to a self-inflationary
solution which produces a late accelerated expansion such that the
radius becomes a function of $n$.
\end{abstract}
\vspace{2cm}

\section{Introduction}
The notion of extra dimensions and that they can be probed by
gravitons and eventually non-standard matter has been the focus of
attention in recent years. These models usually yield the correct
Newtonian $(1/r)$-potential at large distances because the
gravitational field is quenched on sub-millimeter transverse
scales. This quenching appears either due to finite extension of
the transverse dimensions \cite{1,2} or due to sub-millimeter
transverse curvature scales induced by negative cosmological
constants \cite{3,4,5,6,7,8}. A common feature of both of these
types of models and also of the old Kaluza-Klein type models is
the prediction of deviations from four-dimensional Einstein
gravity at short distances. If the transverse length scale is not
too small, this implies the possibility to generate bulk gravitons
in accelerators or stars \cite{2,9}. The model of Dvali, Gabadadze
and Porrati (DGP) \cite{10}, see \cite{11} for extensions, is
different, predicting that 4-dimensional Einstein gravity is a
short-distance phenomenon with deviations showing up at large
distances. The transition between four and higher-dimensional
gravitational potentials in the DGP model arises as a consequence
of the presence of both the brane and bulk Einstein terms in the
action. Further, it was observed in \cite{13,14} that the DGP
model allows for an embedding of the standard Friedmann cosmology
in the sense that the cosmological evolution of the background
metric on the brane can entirely be described by the standard
Friedmann equation plus energy conservation on the brane. This was
later generalized to arbitrary number of transverse dimensions in
\cite{15}. For a recent and comprehensive review of the
phenomenology of DGP cosmology, the reader is referred to
\cite{lue}.

An interesting observation made a few years ago is that the
expansion of our universe is currently undergoing a period of
acceleration which is directly measured from the light-curves of
several hundred type Ia supernovae \cite{16,17} and independently
from observations of the cosmic microwave background (CMB) by the
WMAP satellite \cite{18} and other CMB experiments \cite{19}.
However, the mechanism responsible for this acceleration is not
well understood and many authors introduce a mysterious cosmic
fluid, the so called dark energy, to explain this effect
\cite{20}. Recently, it has been shown that such an accelerated
expansion could be the result of a modification to the
Einstein-Hilbert action \cite{21}. A scenario where the issue of
cosmic acceleration in the framework of higher order theories of
gravity in $4D$ is addressed can be found in \cite{capo}. One of
the first proposals in this regards was suggested in \cite{22}
where a term of the form $R^{-1}$ was added to the usual
Einstein-Hilbert action. It was then shown that this term could
give rise to accelerating solutions of the field equations without
dark energy. In \cite{ata} a DGP brane model with a scalar field
on the brane was proposed, predicting that for the gravitational
potential, the mass density should be modified by the addition of
the mass density of the scalar filed on the brane. Such a scalar
field has its origin in the conformal transformation used to
transform the action in the Jordan frame to the Einstein frame in
the usual $4D$ modified gravity. For non-minimally coupled scalar
field scenarios see \cite{nozari}.

In this paper, we focus attention on the DGP brane model where the
action contains an arbitrary function of the Ricci scalar, ${\cal
L}(R)$, and obtain the evolution of the metric on the space-time.
We concentrate on a specific form for ${\cal L}(R)={\cal
L}_{0}R^n$ and solve the resulting dynamical equations, predicting
a power-law acceleration on the brane. The components of the
metric in the Gaussian normal coordinates are then calculated and
presented. Finally, we show that there exists two classes of
solutions close to the usual FRW cosmology for small Hubble radii
in the model presented here.
\section{DGP model with ${\cal L}(R)$ brane action}
We start by writing the action for the DGP model with an arbitrary
function of the scalar curvature on the brane part of the action,
that is
\begin{equation}\label{eq1}
{\cal S}=\frac{m^{3}_{4}}{2}\int d^{5}x\sqrt{-g}{\cal
R}+\frac{m^{2}_{3}}{2}\int d^{4}x\sqrt{-q}{\cal L}(R)+{\cal
S}_{m}\left[q_{\mu\nu},\psi_{m}\right],
\end{equation}
where the first term in (\ref{eq1}) corresponds to the
Einstein-Hilbert action in $5D$ for the 5-dimensional bulk metric
$g_{AB}$, with the Ricci scalar denoted by ${\cal R}$. Similarly,
the second term is the modified Einstein-Hilbert action with
${\cal L}(R)$ corresponding to the induced metric $q_{\mu\nu}$ on
the brane, where ${\cal L}(R)$ is some arbitrary function of the
$4D$  scalar curvature and $m_{3}$ and $m_{4}$ are reduced Planck
masses in four and five dimensions respectively with ${\cal
S}_{m}$ being the matter action on the brane with a matter field
denoted by $\psi_{m}$. The induced metric $q_{\mu\nu}$ is defined
as usual from the bulk metric $g_{AB}$ by
\begin{equation}\label{eq2}
q_{\mu\nu}=\delta^{A}_{\mu}\delta^{B}_{\nu}g_{AB}.
\end{equation}
It would now be possible to write the field equations resulting from
this action, yielding, in $d-1$ spatial dimensions
\begin{equation}\label{eq3}
\frac{m^{3}_{4}}{{\cal L}'(R)}\left({\cal
R}_{AB}-\frac{1}{2}g_{AB}{\cal
R}\right)+m^{2}_{3}\delta^{\mu}_{A}\delta^{\nu}_{B}\left(R^{(d-1)}_{\mu\nu}-
\frac{1}{2}q_{\mu\nu}R^{(d-1)}\right)\delta(y)
=\delta^{\mu}_{A}\delta^{\nu}_{B}\left(\hat{T}_{\mu\nu}+
T^{(\it{curv})}_{\mu\nu}\right)\delta(y),
\end{equation}
where $\hat{T}_{\mu\nu}=\frac{1}{{\cal L}'(R)}T_{\mu\nu}$ and
$T_{\mu\nu}$ is the energy-momentum tensor in the matter frame,
and
\begin{equation}\label{eq4}
T^{({\it curv })}_{\mu\nu}=\frac{m^{2}_{3}}{{\cal
L'}(R)}\left\{\frac{1}{2}q_{\mu\nu}[{\cal L}(R)-R{\cal L'}(R)]+{\cal
L'}(R)^{;\alpha\beta}(q_{\mu\alpha}q_{\nu\beta}-q_{\mu\nu}q_{\alpha\beta})\right\}.
\end{equation}
A prime here denotes differentiation with respect to $R$. Note
that $R$ lives on the brane. The corresponding junction
conditions, relating the extrinsic curvature to the
energy-momentum tensor, become
\begin{eqnarray}\label{eq6}
\mbox{lim}_{\epsilon\rightarrow+0}[K_{\mu\nu}]^{y=+\epsilon}_{y=-\epsilon}
&=&\left.\frac{{\cal L}'(R)}{m^{3}_{4}}\left(\hat{T}_{\mu\nu}+
T^{({\it
curv})}_{\mu\nu}-\frac{1}{d-1}q_{\mu\nu}q^{\alpha\beta}\left(\hat{T}_{\alpha\beta}+
T^{({\it curv})}_{\alpha\beta}\right)\right)\right|_{y=0}\nonumber\\
&-&\left.\frac{{\cal
L}'(R)m^{2}_{3}}{m^{3}_{4}}\left(R^{(d-1)}_{\mu\nu}-
\frac{1}{2(d-1)}q_{\mu\nu}q^{\alpha\beta}R^{(d-1)}_{\alpha\beta}\right)\right|_{y=0}.
\end{eqnarray}
It should be noted that ${\cal L}'(R)$ is a function of the brane
parameters.
\section{Cosmology }

From the fact that the DGP model predicts deviations only at large
distances, one might hope it could be ruled out from cosmological
observations. However, as we shall see, it could account for
cosmological equations of motion at any distance scale on the
brane with any function of the Ricci scalar. Brane cosmology
usually starts from the line element
\begin{equation}\label{eq15}
ds^{2}=q_{\mu\nu}dx^{\mu}dx^{\nu}+b^{2}(y,t)dy^{2}=-n^{2}(y,t)dt^{2}+
a^{2}(y,t)\gamma_{ij}dx^{i}dx^{j}+b^{2}(y,t)dy^{2},
\end{equation}
where $\gamma_{ij}$ is a maximally symmetric 3-dimensional metric
with $k=-1,0,1$ being the usual parameters denoting the spatial
curvatures. Building on the results of \cite{26,27}, the
cosmological evolution equations of a 3-brane in a $5D$ bulk
resulting from equations (\ref{eq3}) and (\ref{eq6}) were
presented in the first two references in \cite{21}. Here we will
follow \cite{13} and give the results for a brane of dimension
$\nu+1$. Adopting the Gaussian normal system gauge
\begin{equation}\label{eq16}
b^{2}(y,t)=1,
\end{equation}
The field equations on the brane for metric (\ref{eq15}) and $d
=\nu+1$ spatial dimensions are
\begin{equation}\label{eq17}
G^{(\nu)}_{00}=\frac{1}{2}\nu(\nu-1)n^{2}\left(\frac{\dot{a}^{2}}{n^{2}a^{2}}+
\frac{k}{a^{2}}\right),
\end{equation}
\begin{equation}\label{eq18}
G^{(\nu)}_{ij}=(\nu-1)\left(\frac{\dot{n}\dot{a}}{n^{3}a}-
\frac{\ddot{a}}{n^{2}a}\right)q_{ij}-\frac{1}{2}(\nu-1)(\nu-2)
n^{2}\left(\frac{\dot{a}^{2}}{n^{2}a^{2}}+
\frac{k}{a^{2}}\right)q_{ij}.
\end{equation}
The junction conditions (\ref{eq6}) for an ideal fluid on the brane,
given by
\begin{equation}\label{eq24}
T_{\mu\nu}=(\rho+p)u_{\mu}u_{\nu}+pq_{\mu\nu},
\end{equation}
read
\begin{eqnarray}\label{eq25}
\lim_{\epsilon\rightarrow+0}\left[\partial_{y}n\right]^{y=+\epsilon}_{y=-\epsilon}=\frac{n{\cal
L'}(R)}{\nu
m^{\nu}_{\nu+1}}\left[(\nu-1)\rho^{({\it tot})}\left.+\nu p^{({\it tot})}\right]\right|_{y=0}\nonumber\\
+\frac{{\cal
L'}(R)m^{\nu-1}_{\nu}}{m^{\nu}_{\nu+1}}(\nu-1)n\left[-\frac{\dot{n}\dot{a}}{n^{3}a}+
\frac{\ddot{a}}{n^{2}a} \left.-\frac{\dot{a}^{2}}{2n^{2}a^{2}}-
\frac{k}{2a^{2}}\right]\right|_{y=0},
\end{eqnarray}
\begin{equation}\label{eq26}
\lim_{\epsilon\rightarrow+0}\left[\partial_{y}a\right]^{y=+\epsilon}_{y=-\epsilon}=
\left.\frac{{\cal
L'}(R)m^{\nu-1}_{\nu}}{2m^{\nu}_{\nu+1}}(\nu-1)\left[\frac{\dot{a}^{2}}{n^{2}a}+
\frac{k}{a}\right]\right|_{y=0} -\left.\frac{{\cal L'}(R)\rho^{({\it
tot})}a}{\nu m^{\nu}_{\nu+1}}\right|_{y=0},
\end{equation}
where
\begin{equation}\label{eq27}
\rho^{({\it tot})}=\hat{\rho}+\rho^{({\it curv})},
\end{equation}
\begin{equation}\label{eq28}
p^{({\it tot})}=\hat{p}+p^{({\it curv})},
\end{equation}
and $\hat{\rho}$ and $\hat{p}$ are the energy density and pressure
in the matter frame associated with $\hat{T}_{\mu\nu}$
respectively. Energy conservation on the brane follows from the
vanishing of $\frac{1}{{\cal L'}(R)}G_{05}=0 \Rightarrow
\nu\left(\frac{n'}{n}\frac{\dot{a}}{a}-\frac{\dot{a}'}{a}\right)=0$.
We obtain
\begin{equation}\label{eq29}
\frac{n'}{n}=\frac{\dot{a}'}{\dot{a}},
\end{equation}
and in particular
\begin{equation}\label{eq30}
\lim_{\epsilon\rightarrow+0}\left[\frac{n'}{n}\right]^{y=+\epsilon}_{y=-\epsilon}=
\lim_{\epsilon\rightarrow+0}\left[\frac{\dot{a}'}{\dot{a}}\right]^{y=+\epsilon}_{y=-\epsilon}.
\end{equation}
Insertion of (\ref{eq25}) and (\ref{eq26}) into this equation yields
the equation of conservation
\begin{equation}\label{eq31}
\left.\dot{\rho}^{({\it tot})}a\right|_{y=0}=
-\nu\left.\left(\rho^{({\it tot})}+p^{({\it
tot})}\right)\dot{a}\right|_{y=0}.
\end{equation}
Also, insertion of (\ref{eq29}) into the bulk equations $G_{00}$
and $G_{55}$ for $y\neq0$ yields a $\nu$-dimensional version of
the integral of \cite{27}, that is
\begin{equation}\label{eq32}
\frac{1}{{\cal L'}(R)}G_{00}=0\Rightarrow\frac{2}{\nu
n^{2}}a'a^{\nu}G_{00}=\frac{\partial}{\partial y}I=0,
\end{equation}
and
\begin{equation}\label{eq33}
\frac{1}{{\cal
L'}(R)}G_{55}=0\Rightarrow\frac{2}{\nu}\dot{a}a^{\nu}G_{55}=-\frac{\partial}{\partial
t}I=0.
\end{equation}
This means that if we define the quantities $I^+$ and $I^-$ by
taking the factor $a^{\nu-1}$ out of the right hand sides of
equations (\ref{eq32}) and (\ref{eq33}), that is
\begin{equation}\label{eq34}
I^{+}=\left.\left(\frac{\dot{a}^{2}}{n^{2}}-a'^{2}+k\right)a^{\nu-1}\right|_{y>0},
\end{equation}
\begin{equation}\label{eq35}
I^{-}=\left.\left(\frac{\dot{a}^{2}}{n^{2}}-a'^{2}+k\right)a^{\nu-1}\right|_{y<0},
\end{equation}
then $I^+$ and $I^-$ are constants with the property that $I^+ =
I^-$ if
\begin{equation}\label{eq36}
\lim_{\epsilon\rightarrow+0}a'|_{y=+\epsilon}=\pm
\lim_{\epsilon\rightarrow+0}a'|_{y=-\epsilon}.
\end{equation}
We can now simplify the previous equations by further restricting
the gauge
\begin{equation}\label{eq37}
n(0,t)=1,
\end{equation}
and by simply performing the transformation
\begin{equation}\label{eq38}
t=\int^{t}n(0,\tau)d\tau,
\end{equation}
of the time coordinate. This gauge is convenient because it gives
the usual cosmological time on the brane. Using equations
(\ref{eq29}) and (\ref{eq37}), we find that our basic dynamical
variable is $a(y,t)$ with $n(y,t)$ given by
\begin{equation}\label{eq39}
n(y,t)=\frac{\dot{a}(y,t)}{\dot{a}(0,t)}.
\end{equation}
The basic set of cosmological equations in the present setting for
any function of the Ricci scalar on the brane in the DGP model
without a cosmological constant in the bulk are thus equations
(\ref{eq26}), (\ref{eq31}), (\ref{eq34}) and (\ref{eq35}) which
have to be amended with dispersion relations (or the corresponding
evolution equations) for the ideal fluid components on the brane,
that is
\begin{eqnarray}\label{eq40}
\lim_{\epsilon\rightarrow+0}\left[\partial_{y}a\right]^{y=
+\epsilon}_{y=-\epsilon}(t)&=&\frac{{\cal
L'}(R)m^{\nu-1}_{\nu}}{2m^{\nu}_{\nu+1}}(\nu-1)
\left[\frac{\dot{a}^{2}(0,t)}{a(0,t)}+
\left.\frac{k}{a(0,t)}\right]\right|_{y=0} \nonumber\\
&-&\left.\frac{{\cal L'}(R)(\hat{\rho}+\rho^{({\it
curv})})a(0,t)}{\nu m^{\nu}_{\nu+1}}\right|_{y=0},
\end{eqnarray}
\begin{equation}\label{eq41}
I^{+}=\left.\left[\dot{a}^{2}(0,t)-a'^{2}(y,t)+k\right]a^{\nu-1}(y,t)\right|_{y>0},
\end{equation}
\begin{equation}\label{eq42}
I^{-}=\left.\left[\dot{a}^{2}(0,t)-a'^{2}(y,t)+k\right]a^{\nu-1}(y,t)\right|_{y<0},
\end{equation}
\begin{equation}\label{eq43}
\rho^{({\it curv})}=\frac{m^{2}_{3}}{{\cal
L'}(R)}\left\{\frac{1}{2}[{\cal L}(R)-R{\cal
L'}(R)]-3\left(\frac{\dot{a}(0,t)}{a(0,t)}\right)\dot{R}{\cal
L''}(R)\right\},
\end{equation}
\begin{equation}\label{eqq43}
p^{({\it curv})}=\frac{m^{2}_{3}}{{\cal
L'}(R)}\left\{2\left(\frac{\dot{a}(0,t)}{a(0,t)}\right)\dot{R}{\cal
L''}(R)+\ddot{R}{\cal L''}(R)+\dot{R}^{2}{\cal
L'''}(R)-\frac{1}{2}[{\cal L}(R)-R{\cal L'}(R)]\right\},
\end{equation}
\begin{equation}\label{eq44}
n(y,t)=\frac{\dot{a}(y,t)}{\dot{a}(0,t)}.
\end{equation}
Let us now discuss the cosmology in the DGP model by taking
$I^{+}=I^{-}$. The cosmological equations in this framework for a
$(\nu-1)$-dimensional space are given by
\begin{equation}\label{eq45}
\frac{\dot{a}^{2}(0,t)+k}{a^{2}(0,t)}=
\frac{2(\hat{\rho}+\rho^{({\it curv})})}{\nu(\nu-1)m^{\nu-1}_{\nu}},
\end{equation}
\begin{equation}\label{eq47}
I=\left[\dot{a}^{2}(0,t)-a'^{2}(y,t)+k\right]a^{\nu-1}(y,t),
\end{equation}
\begin{equation}\label{eq48}
n(y,t)=\frac{\dot{a}(y,t)}{\dot{a}(0,t)}.
\end{equation}
The evolution of the background geometry of the observable universe
according to the Friedmann equation can thus be embedded in the DGP
model, with the behavior of $a(y, t)$ off the brane determined
solely by the integral $I$ and the boundary condition $a(0, t)$ from
the Friedmann equation. This embedding will be asymmetric in all
realistic cases, because the requirement that the Friedmann equation
holds on the brane is equivalent to the smoothness condition
\begin{equation}\label{eq49}
\lim_{\epsilon\rightarrow+0}
a'(\epsilon,t)=\lim_{\epsilon\rightarrow+0} a'(-\epsilon,t).
\end{equation}
This could yield a symmetric embedding only for $a'(0, t)= 0$, but
this is incompatible with the time independence of the integral
$I$ apart from the case $k = -1$, $\dot{a}=1 $. For a discussion
of this point the reader may consult \cite{28,29}. In the present
calculations we will choose the sign of $y$ in the direction of
the increasing scale factor, $a'>0$. The possibility of a direct
embedding of Friedmann cosmology is a consequence of the fact that
the evolution of the background geometry (\ref{eq15}) and the
source terms $\rho^{({\it curv})}$ and $p^{({\it curv})}$ are
supposed to depend only on $t$ and $y$. For $\nu=3$ we obtain
\begin{equation}\label{eq50}
I=\left[\dot{a}^{2}(0,t)-a'^{2}(y,t)+k\right]a^{2}(y,t),
\end{equation}
and from the equation for $n(y, t)$, the solutions for the metric
components off the brane in terms of the metric on the brane
(assuming $a'>0$) are
\begin{equation}\label{eq51}
a^{2}(y,t)=a^{2}(0,t)+
(\dot{a}^{2}(0,t)+k)y^{2}+2\sqrt{(\dot{a}^{2}(0,t)+k)
a^{2}(0,t)-I}y,
\end{equation}
\begin{equation}\label{eq52}
n(y,t)=\left[a(0,t)+ \ddot{a}(0,t)y^{2}+a(0,t)y
\frac{a(0,t)\ddot{a}(0,t)+
\dot{a}^{2}(0,t)+k}{\sqrt{(\dot{a}^{2}(0,t)+k)
a^{2}(0,t)-I}}\right]\frac{1}{a(y,t)}.
\end{equation}
This embedding of the Friedmann cosmology on the brane becomes
particularly simple for $I = 0$, that is
\begin{equation}\label{eq53}
a(y,t)=a(0,t)+\sqrt{\dot{a}^{2}(0,t)+k}y,
\end{equation}
\begin{equation}\label{eq54}
n(y,t)=1+\frac{\ddot{a}(0,t)}{\sqrt{\dot{a}^{2}(0,t)+k}}y.
\end{equation}

\section{$R^{n}$ gravity in the DGP model}

To progress further, the form of ${\cal L}(R)$ should be
specified. For ease of exposition and clarity, let us focus
attention on theories where a $R^{n}$ term is present in the
action and write
\begin{equation}\label{eq55}
{\cal L}(R)={\cal L}_{0}R^{n}.
\end{equation}
Let us also assume a power law solution form for the brane scale
factor as
\begin{equation}\label{eqq55}
a(0,t)=a_{0}\left(\frac{t}{t_{0}}\right)^{\alpha}.
\end{equation}
The interesting cases are for $\alpha\geq1$ which give rise to
acceleration.

To proceed, we consider the evolution of the scale factor with
time on the brane. Using equation (\ref{eq45}) for the spatially
flat FRW metric and setting $\nu=3$, we write
\begin{equation}\label{eq56}
\left(\frac{\dot{a}(0,t)}{a(0,t)}\right)^{2}=\frac{1}{3m^{2}_{3}}\rho^{({\it
tot})}.
\end{equation}
Now, use of the conservation equation leads to
\begin{equation}\label{eq57}
\frac{\ddot{a}(0,t)}{a(0,t)}=-\frac{1}{6m^{2}_{3}}\left[\rho^{({\it
tot})}+3p^{({\it tot})}\right].
\end{equation}
We must now solve the system of equations (\ref{eq56}) and
(\ref{eq57}) with $\hat{\rho}=\hat{p}=0$. Substituting equations
(\ref{eq55}) and (\ref{eqq55}) into the above equations describing
the dynamical system for the brane, we obtain an algebraic system
for the parameters $n$ and $\alpha$
\begin{eqnarray}\label{eq59}
\alpha[\alpha(n-2)+2n^{2}-3n+1]=0\nonumber, \\
\alpha[\alpha(n-2)+(n^{2}-n+1)]=n(2n-1)(n-1),
\end{eqnarray}
from which the allowed solutions
\begin{eqnarray}\label{eq60}
\alpha=0\Rightarrow\,\,\,\ n=0,\,\,1/2,\,\,\,1,\nonumber \\
\alpha=\frac{2n^{2}-3n+1}{2-n},\,\,\forall\,\, n
\,\,\,\mbox{except} \,\,\ n=2,
\end{eqnarray}
follow. The solutions with $\alpha=0$ are not interesting since
they provide static cosmologies with a non-evolving scale factor
on the brane, matching cosmological models resulting from the
solutions of the Einstein equations without matter and $n=1$. This
particular value of $n$ warrants a further discussion. It is well
known that in ordinary $4D$ gravity an action of the form ${\cal
L}(R)={\cal L}_{0}R^{n}$ results in a constant scale factor in a
cosmological setting without any ordinary matter and a singular
equation of state for $n=1$ \cite{capo}.  It is therefore not
surprising to expect the same behavior in DGP models when the
action on the brane is taken as that mentioned above. However, if
one takes ${\cal L}(R)=R+{\cal L}_{0}R^{n}$, one is lead to
ordinary general relativity as the low energy limit of the theory
and one finds that this theory is equivalent to scalar
quintessence models through conformal transformations \cite{22}.
The corresponding result for ${\cal L}(R)={\cal L}_{0}R^{n}$, $n >
0$ and $ n\neq 1$ is also spelled out in \cite{capo}. Using
equations (\ref{eq43}) and (\ref{eqq43}) we can deduce the
equation of state for the family of solutions $\alpha\neq0$. We
then have
\begin{eqnarray}\label{eqq60}
w^{({\it curv})}=-\left(\frac{6n^{2}-7n-1}{6n^{2}-9n+3}\right),
\end{eqnarray}
which clearly behaves as $w^{({\it curv})}\rightarrow -1$ for
$n\rightarrow\infty$, playing the role of a cosmological constant.
For $R^{-1}(n=-1)$ which corresponds to $\alpha=2$ we have a
power-law acceleration on the brane without having to introduce
dark energy. This result is consistent with the observational
results similar to dark energy with the equation of state
parameter $-1<w^{({\it curv})}=\frac{-2}{3}<-\frac{1}{3}$
\cite{30}. Using equations (\ref{eqq55}) and (\ref{eq60}) we
obtain the deceleration parameter
\begin{eqnarray}\label{eqqq60}
q(n)=-\left(\frac{2n^{2}-2n-1}{2n^{2}-3n+1}\right).
\end{eqnarray}
It is clear that for $n\rightarrow \infty$ we have $q(\infty)=-1$.
This means that the universe (brane) is continuously expanding at
an ever-increasing rate. Clearly, for $n=-1$ or $\alpha=2$ we
obtain $q=-\frac{1}{2}$. It is worth noting again, as was
mentioned before, that in the normal DGP model, that is when $n=1$
and with no ordinary matter present, we cannot define the equation
of state and deceleration parameter because the scale factor on
the brane is constant, see equations (\ref{eq56}) and
(\ref{eq57}). We therefore expect the same behavior in our model
and see that these equations diverge for $n=1$. Figure 1 shows the
behavior of $q$ as a function of $n$. As can be seen, for $n<-0.4$
and $n>1.4$, $q\rightarrow -1$.
\begin{figure}
\begin{center}
\epsfig{figure=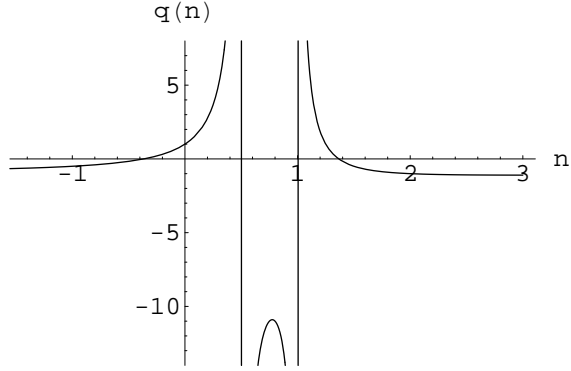,width=8cm}\hspace{5mm}
\end{center}
\caption{\footnotesize Behavior of $q(n)$ as a function of $n$.}
\end{figure}

Let us now find the evolution of $a(y,t)$ and $n(y,t)$ everywhere
in space-time. Substituting equations (\ref{eqq55}) and
(\ref{eq60}) into equations (\ref{eq51}) and (\ref{eq52}), one
finds
\begin{equation}\label{eq61}
a^{2}(y,t)=a_{0}^{2}\left(\frac{t}{t_{0}}\right)^{2\alpha}\left[1+\frac{\alpha^{2}}{\left(\frac{t}{t_{0}}\right)^{2}}y^{2}\right]+
2\sqrt{a_{0}^{4}\alpha^{2}\left(\frac{t}{t_{0}}\right)^{4\alpha}-I}y,
\end{equation}
and
\begin{equation}\label{eq62}
n(y,t)=a_{0}\left(\frac{t}{t_{0}}\right)^{\alpha}\left[1+\frac{\alpha(\alpha-1)}{\left(\frac{t}{t_{0}}\right)^{2}}y^{2}+
a_{0}\left(\frac{t}{t_{0}}\right)^{2\alpha-2}\frac{2\alpha^{2}-\alpha}{\sqrt{a_{0}^{4}\alpha^{2}\left(\frac{t}{t_{0}}\right)^{4\alpha-2}-I}}y
\right]\frac{1}{a(y,t)}.
\end{equation}
In the particular case $I=0$, we obtain
\begin{equation}\label{eq63}
a(y,t)=a_{0}\left(\frac{t}{t_{0}}\right)^{\alpha}\left[1+\frac{\alpha}{\left(\frac{t}{t_{0}}\right)}y\right],
\end{equation}
and
\begin{equation}\label{eq64}
n(y,t)=\left[1+\frac{(\alpha-1)}{\left(\frac{t}{t_{0}}\right)}y\right].
\end{equation}
The scale factor on the full space-time, $a(y,t)$, is plotted in
figure 2 for the special case $I=0$ and $\alpha=2$ or $n=-1$. Note
that for $y=0$, equations (\ref{eq63}) and (\ref{eq64}) reduce to
(\ref{eqq55}) and $n(0,t)=1$ respectively. There appears
coordinate singularities on the space-like hypercone $ y=\pm(
\frac{t/t_{0}}{\alpha-1})$. This is presumably a consequence of
the fact that the orthogonal geodesics emerging from the brane
(which we used to set up our Gaussian normal system, $b^{2}=1$) do
not cover the full five-dimensional space-time.
\begin{figure}
\begin{center}
\epsfig{figure=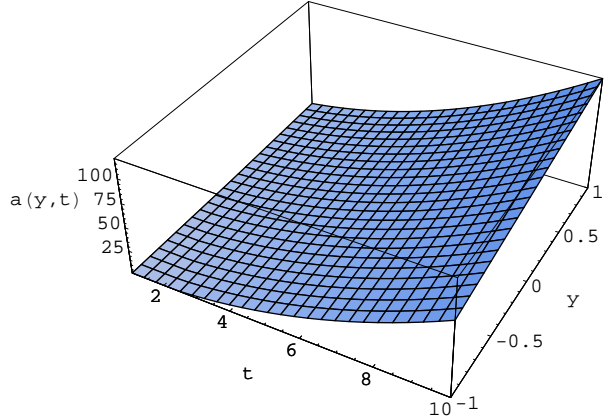,width=8cm}\hspace{5mm}
\end{center}
\caption{\footnotesize Behavior of $a(y,t)$ as a function of $t$
and $y$ for $n=-1$.}
\end{figure}

\section{Brane Friedman equations with ${\cal L}(R)$}
In this section we derive the Friedman equations for the brane
metric in the presence of ${\cal L}(R)$ in the brane part of DGP
action. Thus, for $I=0$ in equation (\ref{eq50}) and using
equation (\ref{eq37}) we can write
\begin{equation}\label{eq65}
\left[\dot{a}^{2}(0,t)-a'^{2}(y,t)\right]a^{2}(y,t)=0.
\end{equation}
Note that we assume $k=0$ and the bulk cosmological constant is
zero. Taking the symmetry $y\leftrightarrow-y$ $(I^{+}=I^{-})$ for
simplicity, equation (\ref{eq40}) can be used to compute $a'(y,t)$
on the two sides of the brane. We have in this case
$[a'(y,t)]=2a'(0^{+},t)$. By continuity when $y\rightarrow0$,
equation (\ref{eq65}) yields the generalized (first) Friedman
equation
\begin{equation}\label{eq66}
H^{2}-2\epsilon\frac{2\mu^{2}}{\kappa^{2}{\cal
L}'(R)}H=\frac{\mu^{2}}{3}\rho^{({\it tot})},
\end{equation}
where the Hubble parameter is $H=\frac{\dot{a}(0,t)}{a(0,t)n(0,t)}$,
$\kappa^{2}=m^{-3}_{5}$, $\mu^{2}=m^{-2}_{4}$ and $\epsilon=\pm1$ is
the sign of $[a'(y,t)]$.

Now, let us discuss the solutions of the Friedman equation
(\ref{eq66}) together with (\ref{eq31}) when the bulk cosmological
constant vanishes. It is now apparent from (\ref{eq66}) that the
standard cosmology, namely the usual $4D$ Friedman equation
(\ref{eq56}) is recovered whenever the last term on the left hand
side of (\ref{eq66}) is subdominant with respect to the first
term, namely when
\begin{equation}\label{eq68}
H\gg2\frac{\mu^{2}}{\kappa^{2}{\cal L}'(R)}.
\end{equation}
Thus using ansatzs (\ref{eq55}) and (\ref{eqq55}), the above
equation in terms of the Hubble radius $H^{-1}$ can be written
\begin{equation}\label{eq69}
H^{-1}\ll(r_{c}n{\cal
L}_{0})^{\frac{1}{2n-1}}(12-6/\alpha)^{\frac{n-1}{2n-1}},
\end{equation}
where $r_{c}=\frac{m^{2}_{4}}{2m^{3}_{5}}$. For $n=1$ this matches
the scale $r_{c}$ found in \cite{10}, setting the crossover
between the $4D$ and $5D$ gravity regimes. For the special case
$n=-1$, or $\alpha=2$ for which a power-law acceleration
consistent with $t^{2}$ is obtained we have
\begin{equation}\label{eq70}
H^{-1}\ll(81)^{\frac{1}{3}}(-r_{c}{\cal L}_{0})^{\frac{-1}{3}}.
\end{equation}
Also, for $n\rightarrow2$ $(\alpha\rightarrow\infty)$ one has
\begin{equation}\label{eq71}
H^{-1}\ll(24r_{c}{\cal L}_{0})^{\frac{1}{3}}.
\end{equation}
Thus, the Hubble radius $H^{-1}$ in this model depends on $n$. For
the Hubble radius when equation (\ref{eq68}) is not satisfied,
there are two distinct behaviors depending on $\epsilon$. Equation
(\ref{eq66}) can indeed be rewritten as
\begin{equation}\label{eq72}
H=\epsilon\frac{\mu^{2}}{\kappa^{2}{\cal
L}'(R)}\pm\sqrt{\frac{\mu^{2}}{3}\rho^{({\it
tot})}+\frac{\mu^{4}}{\kappa^{4}{\cal L}'(R)^{2}}}.
\end{equation}
Let us start by examining the case where $\epsilon=-1$. Using
condition $H{\cal L}'(R)\ll r^{-1}_{c}$, one may then expand
(\ref{eq66}) to obtain
\begin{equation}\label{eq73}
\frac{H}{{\cal L}'(R)}=\frac{\kappa^{2}}{6}\rho^{({\it tot})},
\end{equation}
or
\begin{equation}\label{eq74}
H^{2}{\cal L}'(R)^{-2}=\frac{\kappa^{4}}{36}\rho^{({\it tot})^{2}},
\end{equation}
which is the full $5D$ regime with ${\cal L}(R)$ gravity on the
brane. Equation (\ref{eq74}) for $R^{n}$ becomes
\begin{equation}\label{eqq74}
H^{-4n+6}=\frac{\kappa^{4}}{36}(n{\cal
L}_{0})^{2}\left(12-\frac{6}{\alpha}\right)^{2n-2}\left(\hat{\rho}+\rho^{({\it
curv})}\right)^{2}.
\end{equation}
Note that $\rho^{({\it curv})}$ is a function of $R$. One has thus
a transition from a $4D$ to a $5D$ regime. If $\epsilon=1$,
however, $H$ is always larger than $H_{\rm self}$ given by
\begin{equation}\label{eq75}
\tilde{H}_{\rm self}=\frac{2\mu^{2}}{\kappa^{2}},
\end{equation}
where $\tilde{H}_{\rm self}=H{\cal L}'(R)$. For $R^{n}$ we find
\begin{equation}\label{eq76}
H_{\rm self}=\left(\frac{2\mu^{2}}{\kappa^{2}n{\cal
L}_{0}}\right)^{\frac{1}{2n-1}}(12-6/\alpha)^{\frac{1-n}{2n-1}},
\end{equation}
and the expansion will never enter into a fully $5D$ regime. Thus
for $k=0$ or $k=-1$ the Hubbel parameter is bounded from below by
equation (\ref{eq76}) for $R^{n}$ on the brane.

An interesting feature of the normal DGP models is the existence
of ghost-like excitations \cite{luty,nico,koya}. It would
therefore be interesting to briefly mention recent results
relevant to the present work. In \cite{koya}, the author has
shown, in the context of normal DGP models, that if we introduce a
positive cosmological constant on the brane $(Hr_{c} > 1)$, then
the spin-2 graviton will have a mass in the range $0 < m^{2} <
2H^{2}$ and that there is a normalizable brane fluctuation mode
with mass $m^{2} = 2H^{2}$. Although the brane fluctuation mode is
healthy, the spin-2 graviton has a helicity-0 excitation which is
known as a ghost. If we allow a negative cosmological constant on
the brane, the brane bending mode becomes a ghost for $ 1/2 <
Hr_{c }< 1$. This confirms the results obtained by the boundary
effective action that there exists a scalar ghost mode for $Hr_{c}
> 1/2$. In a self-accelerating universe$ Hr_{c} = 1$, the spin-2
graviton has mass $m^{2} = 2H^{2}$, which is known to be a special
case for massive gravitons in de Sitter spacetime where the
graviton has no helicity-0 excitation and so no ghost. However, in
DGP models, there exists a brane fluctuation mode with the same
mass and there arises a mixing between the brane fluctuation mode
and the spin-2 graviton. Thus, according to this scenario, our
model may be prone to having ghosts and this could be the subject
of a separate investigation.
\section{Conclusions}
Brane models generally provide an interesting extension of our
parameter space for gravitational theories. In this work we have
discussed the DGP  model with an arbitrary function of the Ricci
scalar in the brane part of action. The cosmological evolution of
this model was studied by solving the relevant dynamical
equations. The components of the metric on the space-time was
obtained for the $R^{n}$ term in $4D$ gravity. The evolution of
the universe in such a scenario was shown to be consistent with
the present observations, predicting an accelerated expansion.
Finally we have shown that there exists two classes of solutions,
close to the usual FRW cosmology for small enough Hubble radii for
${\cal L}(R)$ gravity.


\begin{thebibliography}{9}
\bibitem{1}N. Arkani-Hamed, S. Dimopoulos, G. Dvali, Phys. Lett. B 429 (1998)
263, [hep-ph/9803315],\\N. Arkani-Hamed, S. Dimopoulos, G. Dvali,
Phys. Rev. D 59 (1999) 086004, [hep-th/9807344].
\bibitem{2}I. Antoniadis, N. Arkani-Hamed, S. Dimopoulos, G. Dvali, Phys. Lett. B 436 (1998) 257, [hep-ph/9804398].
\bibitem{3}L. Randall, R. Sundrum, Phys. Rev. Lett. 83 (1999) 4690, [hep-th/0906064].
\bibitem{4}W. M\"{u}ck, K. S. Viswanathan, I. V. Volovich, Phys. Rev. D 62
(2000) 105019, [hep-th/0004017].
\bibitem{5}R. Gregory, V. A. Rubakov, S. M. Sibiryakov,
Class. Quan. Grav. 17 (2000) 4437, [hep-th/0003109].
\bibitem{6}I. Ya. Aref'eva, M. G. Ivanov, W. M\"{u}ck, K. S. Viswanathan, I.
V.Volovich, Nucl. Phys. B 590 (2000) 273, [hep-th/0004114].
\bibitem{7}M. Cveti$\check{c}$, M. J. Duff, J.T. Liu, H. Lu, C. N. Pope, K. S. Stelle, Nucl. Phys. B 605
(2001) 141, [peh-th/0011167].
\bibitem{8}B. Abdesselam, N. Mohammedi, Phys. Rev. D 65 (2002) 084018, [hep-th/0110143].
\bibitem{9}E. A. Mirabelli, M. Perelstein, M. E. Peskin, Phys. Rev. Lett.
82 (1999) 2236, [hep-ph/981133],\\ T. Appelquist, H. C. Cheng, B. A.
Dobrescu, Phys. Rev. D 64 (2001) 035002, [hep-ph/0201131],\\ T. G.
Rizzo, Phys. Rev. D 64 (2001) 095010, [hep-ph/0106336],\\ S. Cullen,
M. Perelstein, Phys. Rev. Lett. 83 (1999) 268, [hep-ph/9903422],
\\V. Barger, T. Han, C. Kao, R. J. Zhang, Phys. Lett. B 461 (1999) 34, [hep-ph/9905474], \\
S. Cassisi, V. Castellani, S. Degl'Innocenti, G. Fiorentini, B.
Ricci, Phys. Lett. B 481 (2000) 323, [astro-ph/0002182],\\ M.
Biesiada, B. Malec, Phys. Rev. D 65 (2002) 043008,
[astro-ph/0109545],
\\ S. Hannestad, G. G. Raffelt, Phys. Rev. Lett. 88 (2002) 071301,
[hep-ph/0110067].
\bibitem{10}G. Dvali, G. Gabadadze, M. Porrati, Phys. Lett. B 485 (2000) 208, [hep-th/0005016].
\bibitem{11}G. Dvali, G. Gabadadze, Phys. Rev. D 63 (2001) 065007, [hep-th/0008054],\\
G. Dvali, G. Gabadadze, M. Kolanovi$\acute{c}$, F. Nitti, Phys. Rev.
D 65 (2002) 024031, [hep-th/0106058].
\bibitem{13}R. Dick, Class. Quant. Grav. 18 (2001) R1, [hep-th/0105320].
\bibitem{14}R. Dick, Actaphys. Polon. B 32 (2001) 3669, [hep-th/0110162].
\bibitem{15}R. Cordero, A. Vilenkin, Phys. Rev. D 65 (2002) 083519, [hep-th/0107175].
\bibitem{lue}A. Lue, Phys. Rept. 423 (2006) 1,
[astro-ph/0510068].
\bibitem{16}A. G. Riess {\it et. al.}  [Supernova Search Team Collaboration], Astron. J. 116
(1998) 1006, [astro-ph/9805201].
\bibitem{17}S. Perlmutter {\it et. al.}, Astron. J. 517 (1999) 565, [astro-ph/9812133],\\
D. N. Spergel {\it et. al.}, Astrophys. J. Suppl. 148 (2003) 175,
[astro-ph/0302209].
\bibitem{18}C. L. Bennett {\it et. al.}, Astrophys. J. Suppl. 148 (2003) 1, [astro-ph/0302207].
\bibitem{19}C. B. Netterfiled {\it et. al.}, Astrophys. J. 571  (2002) 604,
[astro-ph/0104460],\\ N. W. Halverson {\it et. al.}, Astrophys. J.
568 (2002)  38, [astro-ph/0104489].
\bibitem{20}S. M. Carroll, Living  Rev. Rel. 4 (2001)
1, [astro-ph/0004075].
\bibitem{21}C. Deffayet, Phys. Lett. B 502 (2001) 199, [hep-th/0010186],\\ C.
Deffayet, G. R. Dvali and G. Gabadadze, Phys. Rev. D 65 (2002)
044023, [astro-ph/0105068], \\ C. Deffayet and S. J. Landau, J.
Raux, M. Zaldarriaga and P. Astier, Phys. Rev. D 66 (2002) 024019,
[astro-ph/0201164],\\ J. S. Alcaniz, Phys. Rev. D 65 (2002) 123514, [astro-ph/0202492],\\
D. Jain, A. Dev and J. S. Alcaniz, Phys. Rev. D 66 (2002) 083511, [astro-ph/0206224],\\
A. Lue, R. Scoccimarro, G. Starkman, Phys. Rev. D 69 (2004)
044005, [astro-ph/0307034].
\bibitem{capo}S. Capozziello, Int. J. Mod. Phys. D 11 (2002) 483, [astro-ph/0201033], \\
S. Capozziello, Int. J. Mod. Phys. D 12 (2003) 1969,
[astro-ph/0307018].
\bibitem{22}S. M. Carroll, V. Duvvuri, M. Trodden, M. Turner, Phys. Rev. D 70 (2004) 043528, [astro-ph/0306438].
\bibitem{ata}K. Atazadeh and H. R. Sepangi, Phys. Lett. B 643 (2006) 76, [gr-qc/0610107].
\bibitem{nozari} K. Nozari, Phys. Lett. B 652 (2007) 159,
[hep-th/0707.0719],\\ K. Nozari, JCAP 09 (2007) 003,
[hep-th/0708.1611].
\bibitem{26}P. Bin$\acute{e}$truy, C. Deffayet, D. Langlois, Nucl. Phys. B 565 (2000) 269, [hep-th/9905012].
\bibitem{27}P. Bin$\acute{e}$truy, C. Deffayet, U. Ellwanger, D. Langlois, Phys.
Lett. B 477 (2000) 285, [hep-th/9910219].
\bibitem{28}P. P. Avelino, C. J. A. P. Martins, Astrophys.J. 565 (2002) 661, [astro-ph/0106274].
\bibitem{29}C. Deffayet, G. Dvali, G. Gabadadze, [astro-ph/0106449].
\bibitem{30}T. Padmanabhan, Phys. Rept. 380 (2003) 235, [hep-th/
0212290].
\bibitem{luty}M. A. Luty, M. Porrati and R. Rattazzi, JHEP 0309
(2003) 029, [hep-th/ 0303116].
\bibitem{nico}A. Nicolis and R. Rattazzi, JHEP 0406 (2004) 059,
[hep-th/0404159].
\bibitem{koya}K. Koyama, Phys.Rev. D 72 (2005)
123511, [hep-th/0503191].
\end{thebibliography}
\end{document}